\documentclass[conference]{IEEEtran}
\IEEEoverridecommandlockouts

\usepackage{cite}
\usepackage{censor}
\usepackage{textcomp}
\usepackage[utf8]{inputenc} 
\usepackage[T1]{fontenc}    
\usepackage{hyperref}       
\usepackage{url}            
\usepackage{booktabs}       
\usepackage{amsfonts}       
\usepackage{nicefrac}       
\usepackage{microtype}      
\usepackage{xcolor}         
\usepackage{amsmath,amssymb,amsfonts}
\usepackage{graphicx}
\usepackage{blindtext}
\usepackage{textcomp}
\usepackage[notransparent]{svg}
\usepackage{verbatim}
\usepackage{color,soul}
\usepackage{graphics}
\usepackage{listings}
\usepackage{multirow}
\usepackage{tabularx}
\usepackage{amssymb}
\usepackage{mwe}%
\usepackage{todonotes}
\usepackage{tikz}
\usetikzlibrary{arrows.meta, positioning, shapes.multipart, fit, calc}
\usepackage{algorithm}
\usepackage{algpseudocode}
\usepackage{pgfplots}
\usepackage{siunitx}
\pgfplotsset{compat=newest}
\usepgfplotslibrary{groupplots}
\definecolor{darkred}{rgb}{0.55,0.0,0.0}
\pgfplotsset{compat=1.18}  

\usepackage{url}
\usepackage{hyperref}

\def\BibTeX{{\rm B\kern-.05em{\sc i\kern-.025em b}\kern-.08em
    T\kern-.1667em\lower.7ex\hbox{E}\kern-.125emX}}

\usepackage{xcolor}
\usepackage{xspace}
\newif\ifdraft
\drafttrue

\ifdraft
  \newcommand{\jhanote}[1]{{\textcolor{red}{ ***Shantenu: #1 }}\xspace}
\else
 \newcommand{\jhanote}[1]{}
 \fi

\def\BibTeX{{\rm B\kern-.05em{\sc i\kern-.025em b}\kern-.08em
    T\kern-.1667em\lower.7ex\hbox{E}\kern-.125emX}}
\begin{document}

\title{AAFLOW: Scalable Patterns for Agentic AI Workflows\\

}

\author{\IEEEauthorblockN{Arup Kumar Sarker}
\IEEEauthorblockA{\textit{Department of Computer Science,}  \\ \textit{Biocomplexity Institute and Initiative} \\
\textit{University of Virginia}\\
Charlottesville, VA, USA \\
djy8hg@virginia.edu}
\and
\IEEEauthorblockN{Mills Staylor}
\IEEEauthorblockA{\textit{Department of Computer Science,}  \\ \textit{Biocomplexity Institute and Initiative} \\
\textit{University of Virginia}\\
Charlottesville, VA, USA \\
qad5gv@virginia.edu}
\and
\IEEEauthorblockN{Aymen Alsaadi}
\IEEEauthorblockA{\textit{Department of Computer Science,} \\
\textit{Rutgers University}\\
New Brunswick, NJ, USA \\
aymen.alsaadi@rutgers.edu}
\and
\IEEEauthorblockN{Gregor von Laszewski}
\IEEEauthorblockA{\textit{Biocomplexity Institute and Initiative} \\
\textit{University of Virginia}\\
Charlottesville, VA, USA \\
laszewski@gmail.com}
\and
\IEEEauthorblockN{Shantenu Jha}
\IEEEauthorblockA{\textit{Department of Computer Science} \\
\textit{Rutgers University}\\
\textit{Princeton Plasma Physics Laboratory}\\
Princeton, NJ, USA \\
shantenu.jha@rutgers.edu}
\and
\IEEEauthorblockN{Geoffrey Fox}
\IEEEauthorblockA{\textit{Department of Computer Science,}  \\ \textit{Biocomplexity Institute and Initiative} \\
\textit{University of Virginia}\\
Charlottesville, VA, USA \\
vxj6mb@virginia.edu}
}

\maketitle

\begin{abstract}
Agentic workflows in large language model systems integrate retrieval, reasoning, and memory, but existing frameworks suffer from scalability and reproducibility limitations due to fragmented data orchestration, serialization overhead, and non-deterministic execution. Although these frameworks increase flexibility, they don't have a formal execution model that adheres to the principles of high-performance computing. We introduce \textbf{AAFLOW}, a unified distributed runtime that creates communication-efficient execution plans by modeling agentic workflows as an operator abstraction. Using Apache Arrow and Cylon, AAFLOW creates a zero-copy data plane that allows direct interoperability between preprocessing, embedding, and vector retrieval without the need for serialization overhead. To lower coordination costs, it uses resource-deterministic scheduling and asynchronous batching. While retaining comparable LLM generation throughput, experimental results demonstrate up to \textbf{4.64$\times$} pipeline speedup and \textbf{2.8$\times$} gains in embedding and upsert phases. Rather than LLM inference acceleration, these advantages result from enhanced data flow, batching, and communication efficiency.
\end{abstract}


\section{Introduction}

Large language models (LLMs) are increasingly deployed within \emph{agentic workflows}, where multiple components—including retrieval, reasoning, tool invocation, and memory—are dynamically orchestrated to solve complex tasks in scientific computing \cite{arxiv2508}. By adding adaptive control flow and multi-step reasoning to conventional inference pipelines, these workflows enable more expressive and context-aware AI systems. This change, however, highlights a basic systems problem: data infrastructures built for static, predictable workloads are piled on top of dynamic, non-deterministic agent execution. There are two significant drawbacks to the current systems as a result of this mismatch. \textbf{First, data orchestration bottlenecks dominate performance.} Retrieval-Augmented Generation (RAG) pipelines require large-scale preprocessing, including document chunking, embedding generation, indexing, and vector retrieval \cite{ibm25, intelvector}. Significant data transport and node-to-node communication are required for these processes. Due to serialization, object management, and fragmented execution pipelines, existing distributed frameworks like Dask and Spark have significant overhead, especially in communication-bound regimes \cite{arxiv2406, liang2025dataflow, abeykoon2022high}. \textbf{Second, execution non-determinism limits reproducibility and optimization.} Flexible orchestration is introduced by agentic frameworks like LangChain and LangGraph, but LLM-driven decision-making is given execution control \cite{arxiv2508}. In conventional HPC environments, this leads to dynamic execution routes that are challenging to replicate, profile, and optimize \cite{arxiv2509, llamaindex25}.

Existing approaches address these issues separately. While agentic frameworks increase reasoning flexibility without taking underlying data movement and communication costs into account, distributed data systems maximize throughput for static workloads. Scaling these processes sometimes leads to incredibly disjointed architecture solutions, necessitating coordination between multiple services for retrieval, splitting, data storage, and embedding construction. Scaling costs and overall complexity are greatly increased by this \cite{llamaindex25, rishabh25}. Consequently, agentic workflows cannot be mapped onto effective distributed execution models using a single abstraction.

We argue that, similar to relational abstractions in data systems, agentic workflows can be formalized as a composition of operators. Embedding, retrieval, reasoning, memory access, and index updates are examples of agentic operations that can be mapped to well-defined distributed communication patterns, such as broadcast, shuffle-compute, reduction, and embarrassingly parallel execution. This viewpoint allows for a fundamental change: agentic orchestration can be compiled into a predictable execution plan via high-performance communication primitives rather than being treated as a black-box process \cite{arxiv2105}.

In order to achieve this goal, we provide \textbf{AAFLOW}, a unified distributed runtime that serves as an agentic workflow compiler. Using a zero-copy data plane based on Apache Arrow and Cylon, AAFLOW converts high-level agentic processes into communication-efficient execution graphs. AAFLOW allows for scalable and repeatable agentic pipelines by removing serialization overhead and separating logical execution from resource scheduling. AAFLOW introduces three key design principles: \textbf{Operator-driven execution}: Agentic workflows are expressed as a composition of operators mapped to distributed communication patterns. \textbf{Zero-copy data movement}: Apache Arrow enables direct interoperability between data processing, embedding, and retrieval without serialization. \textbf{Resource-deterministic scheduling}: Execution is decoupled from agent logic, enabling predictable and high-concurrency execution.

A multi-tier memory model is inherited partly from the data infrastructure layer, and an agentic orchestration layer in AAFLOW  enables context-aware retrieval and execution continuity. In order to provide adaptive, multi-step reasoning across sessions, the memory layer serves as a bidirectional link between agents and system state, preserving past context and intermediate outcomes. Workloads can be expressed as batch-oriented tasks across transformation, embedding, and indexing stages thanks to AAFLOW's integration of this orchestration with a distributed embedding and ingestion pipeline that uses asynchronous parallel batching and task-based execution to ensure scalability.
This design decreases write amplification in vector databases, eliminates unnecessary vector operations, and permits high-concurrency execution. AAFLOW removes serialization and I/O bottlenecks by combining zero-copy data transfer with communication-efficient scheduling, offering a reliable and repeatable basis for extensive agentic workflows.
This paper makes the following contributions:

\begin{enumerate}
    \item \textbf{Agentic Operator Abstraction}: We introduce a formal abstraction that maps agentic workflow components (embedding, retrieval, reasoning, memory, upsert) to distributed communication patterns, enabling systematic execution on HPC systems.
    
    \item \textbf{Unified Distributed Runtime}: We design AAFLOW, a zero-copy, communication-efficient runtime that integrates distributed data processing, embedding pipelines, and vector retrieval into a single execution model.
    
    \item \textbf{Resource-Deterministic Execution}: We propose an execution model that separates logical agent behavior from resource scheduling, improving reproducibility and enabling high-concurrency batching.
    
    \item \textbf{System-Level Performance Analysis}: We demonstrate that AAFLOW reduces data movement and coordination overhead, achieving up to 4.64$\times$ pipeline speedup and 2.8$\times$ improvements in embedding and upsert stages, while maintaining identical LLM generation throughput.
\end{enumerate}

\subsection{Scope of Evaluation}

We highlight that LLM inference itself is still GPU-bound and comparable across frameworks, and AAFLOW does not speed it up. Rather, the main bottlenecks in large-scale agentic RAG systems are data orchestration, connectivity, and pipeline execution, which may be optimized to improve performance. But we will add and run experiments with multiple LLMs on GPU clusters in the future.

\section{Architectural Design}

Conventional RAG systems are built as loosely connected pipelines that combine the processes of generation, retrieval, embedding, and preprocessing~\cite{NEURIPS2020_6b493230, llamaindex25}. These systems show three basic drawbacks in distributed contexts, notwithstanding their effectiveness at small scales: (1) fragmented data orchestration resulting in high serialization and data movement overhead in distributed settings~\cite{arxiv2406, rishabh25}, (2) non-deterministic execution due to dynamic agent-driven control flow ~\cite{arxiv2508, yao2023react, shinn2023reflexion}, and (3) lack of a unified abstraction for mapping agentic workflows to distributed execution models ~\cite{liang2025dataflow}. The inconsistent generation of embeddings across scattered nodes causes semantic drift in vector indices. \cite{lewis2020retrieval, arxiv2508}. The lack of a standard data abstraction across various data formats also causes additional serialization overhead, which slows down the training and inference stages.

\begin{figure}[htpb]
    \begin{center}
    \includesvg[width=1.0\linewidth]{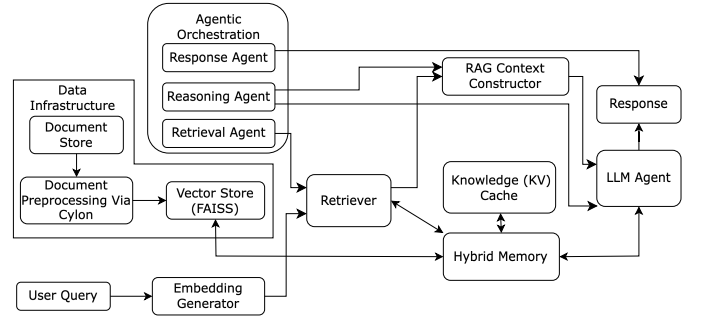}
    \caption{End-to-End code flow of AAFLOW with memory operation. The RAG context constructor uses the memory module through the retriever. } 
    \label{fig:aaflow}
    \end{center}
\end{figure}

AAFLOW proposes a \textbf{formal operator-driven execution model} in place of component-based architectures to overcome these constraints. Agentic workflows are represented in this approach as collections of operators that are assembled into execution graphs over HPC primitives that facilitate communication.

\subsection{Agentic Operator Abstraction}

We define an agentic workflow as a set of operators that modify system state, context, and data. Every operator maps to a particular distributed execution and communication pattern while encapsulating a logical function. This abstraction allows agentic workflows to be compiled into deterministic execution graphs.

Let a workflow be defined as:
\[
\mathcal{W} = \{ Op_{embed}, Op_{retrieve}, Op_{reason}, Op_{memory}, Op_{upsert} \}
\]

Each operator is defined as a tuple:
\[
Op_i = (I_i, O_i, f_i, P_i)
\]
where $I_i$ and $O_i$ denote input and output data, $f_i$ is the transformation function, and $P_i$ is the associated communication pattern.

\vspace{0.5em}

\noindent\textbf{1) Embedding Operator ($Op_{embed}$)}

The embedding operator transforms raw textual data into dense vector representations:
\[
E = Op_{embed}(D)
\]
where $D$ is a set of documents or chunks, and $E$ is the corresponding embedding matrix. Fully parallel execution across nodes is made possible by the separate processing of each document or chunk. It will employ the Embarrassingly Parallel (EP) communication pattern, which makes it simple to divide workloads into smaller, independent subtasks that can be executed concurrently on several cores~\cite{6972327}. The benefits of batch processing and the computational cost are dominated by this operator.

\vspace{0.5em}

\noindent\textbf{2) Retrieval Operator ($Op_{retrieve}$)}

The retrieval operator performs distributed similarity search over partitioned vector indices:
\[
R = Op_{retrieve}(q, E)
\]
where $q$ is the query embedding and $R$ is the set of retrieved top-$k$ documents. Every partition receives the query. Local similarity scores are calculated by each node, and then global aggregation is performed. It employs a communication pattern called shuffle-compute, which is followed by a query broadcast and a partial Top-$k$ reduction between nodes. This operator is dependent on communication and is susceptible to network latency and data partitioning.
\vspace{0.5em}

\noindent\textbf{3) Reasoning Operator ($Op_{reason}$)}

The reasoning operator aggregates retrieved context and produces a structured prompt or response:
\[
C = Op_{reason}(R)
\]
For downstream LLM inference, partial contexts are created locally and merged into a single representation. Distributed context fragments are combined into a single global context using the reduction communication pattern. This operator filters and synthesizes pertinent data to assess the quality of downstream inference.

\vspace{0.5em}

\noindent\textbf{4) Memory Operator ($Op_{memory}$)}

The memory operator manages persistent system state and contextual history:
\[
M' = Op_{memory}(M, C)
\]
Distributed memory stores are used for state updates and lookups. It employs a broadcast/exchange communication pattern in which nodes selectively spread pertinent context or state updates. Adaptive behavior across sessions, context reuse, and multi-turn reasoning are all made possible by this operator.

\vspace{0.5em}

\noindent\textbf{5) Upsert Operator ($Op_{upsert}$)}

The upsert operator inserts or updates embeddings in distributed vector indices:
\[
E' = Op_{upsert}(E)
\]
Updates for embedding are written to partitioned vector stores in batches. In order to preserve index consistency, updates are dispersed and condensed using a shuffle-reduce communication pattern. This operator directly affects ingestion throughput and controls indexing efficiency.

\vspace{0.5em}

AAFLOW is executed as a directed acyclic graph (DAG), where operators are composed as:
\[
Op_{embed} \rightarrow Op_{retrieve} \rightarrow Op_{reason} \rightarrow Op_{memory} \rightarrow Op_{upsert}
\]

While independent operators can be overlapped or parallelized, data dependencies specify the order of execution. This framework allows for explicit communication planning between operators, batching and data location optimization, and deterministic execution independent of LLM control flow. AAFLOW adapts dynamic agentic workflows into predictable distributed programs that may be optimized with HPC scheduling and communication primitives by mapping each operator to a known communication pattern.

\subsection{Compilation to Distributed Execution}

Given a workflow $\mathcal{W}$, AAFLOW constructs a distributed execution plan:

\[
\mathcal{G} = \text{Compile}(\mathcal{W})
\]

where data dependencies are represented by edges and operators are represented by nodes. Every operator is mapped to HPC communication primitives such RDMA-based data exchange, UCX~\cite{shamis2015ucx} transfers, and MPI~\cite{dalcin2005mpi4py} collectives. This stage of compilation guarantees: 1) Deterministic execution pathways that are not dependent on LLM control flow, 2) Explicit communication structures that take the place of implicit framework coordination, 3) Partition-aware scheduling for optimized data locality. Data infrastructure, agentic orchestration, RAG, and a memory module (knowledge cache) are some of the components that make up AAFLOW, which uses distributed determinism and hybrid orchestration shown in Fig.-\ref{fig:aaflow}.

\subsection{Data Infrastructure—Unified Zero-Copy Data Plane}
\begin{figure}[htpb]
    \begin{center}
        \includesvg[width=0.7\linewidth]{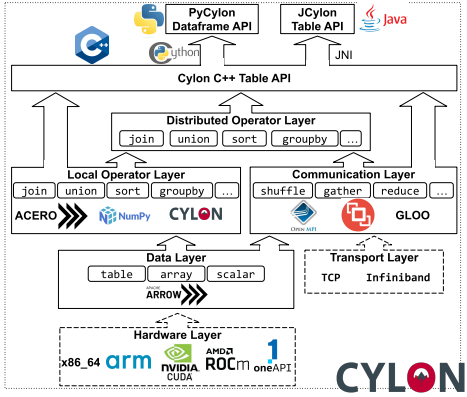}
        \end{center}
        \caption{Cylon Layered Architecture. From the bottom-up view, the Hardware layer is compatible with vendor-based or open-sourced transport layer \cite{perera2023depth}}
    \label{fig:cylon-arch}
\end{figure}

AAFLOW removes serialization overhead between pipeline stages by introducing a shared data plane based on Apache Arrow and Cylon \cite{widanage2020high}. When integrated into hybrid frameworks like as Deep Radical-Cylon (Deep RC) and RHAPSODY \cite{perera2023depth, osti_10634837, alsaadi2025rhapsody}, Cylon provides the high-performance data base required for scalable, repeatable, and low-latency RAG pipelines operating on modern HPC and AI infrastructures. Arrow allows columnar, zero-copy memory sharing between distributed data preprocessing, embedding generation pipelines, and vector database indexing and retrieval, in contrast to conventional frameworks that depend on object-based data transfer. In Fig. \ref{fig:cylon-arch}, cylon's dual-layer interface for Python and C++ guarantees seamless integration with contemporary AI toolchains \cite{alsaadi2025rhapsody}, striking a balance between quick local execution and effective inter-node communication for distributed workflows \cite{cylondoc, sarker2024radical, merzky2022radical}.

Cylon uses network-level communication primitives based on TCP, InfiniBand, and unified backends like MPI\cite{dalcin2005mpi4py}, UCX \cite{shamis2015ucx}, GLOO \cite{gloo:online} and FMI \cite{staylor2025combining} to satisfy the bandwidth and latency requirements of RAG preprocessing (Fig. \ref{fig:cylon-arch}). Data-intensive distributed operations that rely on these communication layers, such as shuffle, gather, and reduce, dominate RAG indexing and embedding generation processes. The resulting abstraction supports high-throughput, communication-aware data movement, which lowers I/O overhead and improves scalability in multi-node situations. \cite{shan2022hybrid, perera2023supercharging}.

\subsection{Resource-Deterministic Execution Model}

The fact that LLM-driven control flow determines execution order, which results in non-deterministic behavior, is a major drawback of current agentic frameworks. This is addressed by AAFLOW by dividing:
\textbf{physical execution}, which involves scheduling and resource allocation, and \textbf{logical execution}, which comprises agent decisions and reasoning stages. Batches of operator executions that are scheduled separately from the agent logic are used to express workloads. This allows for reproducible execution traces, high-concurrency batching, and predictable execution delay. We consider an ingestion workload of $N$ items processed in batches of size $b$ with up to $P$ parallel workers. The cost of processing a batch is modeled as
\begin{equation}
T_{\text{batch}} = \alpha + \beta b,
\end{equation}
where $\alpha$ denotes fixed per-request overhead and $\beta$ denotes per-item cost. With batching and asynchronous execution, the total runtime is approximated as
\begin{equation}
T_{\text{AAFLOW}} \approx \frac{N}{bP}(\alpha + \beta b)
= \frac{N\alpha}{bP} + \frac{N\beta}{P}.
\end{equation}
This formulation shows that parallelism reduces the execution time of the useful work by a factor of $P$, while batching amortizes overhead by a factor of $b$. Across all models, execution time can be expressed as
\begin{equation}
T \approx \frac{N\beta}{P} + \frac{N\alpha}{bP} + \Omega,
\end{equation}
where $\Omega$ represents framework-specific overhead where $\Omega \approx 0$ for ideal async batching. In traditional frameworks (Ray, Dask), $\Omega \approx 0$ includes serialization, task scheduling, and object-store overhead. In contrast, AAFLOW minimizes $\Omega \approx 0$ by 1) eliminating serialization through zero-copy data exchange, 2) reducing coordination via operator-level batching, and 3) using explicit communication primitives instead of implicit task graphs.

\subsection{Agent-Based Retrieval-Augmented Generation}
\label{subsec:agent_rag}


The agents in AAFLOW dynamically decides when to retrieve, how to reformulate queries, and how to integrate intermediate results instead of carrying out a predetermined sequence. As a result, RAG becomes a decision-driven system with the ability to reason iteratively rather than a linear pipeline \cite{yao2023react, shinn2023reflexion}.

Formally, given a user query $q$, the agent performs \textit{query interpretation and planning}, where it decides whether retrieval is required and may decompose $q$ into sub-queries $\{q_1, q_2, \dots, q_n\}$. Each sub-query is embedded into a vector representation and used to retrieve relevant documents from an external knowledge base (e.g., a vector database such as ChromaDB \cite{hong2025context, hong2025benchmarking}, FAISS \cite{johnson2019faiss}). In contrast to static RAG, multi-hop reasoning and better recall are made possible by the agent's ability to iteratively adjust searches depending on intermediate retrieval outcomes (using LlamaIndex agentic orchestration\cite{liu2023llamaindex}). The agent then filters, ranks, and aggregates evidence in a \textit{context integration} stage of processing the retrieved documents. This stage is essential for lowering noise and guaranteeing that the LLM receives only high-quality, task-relevant data (in Fig.-\ref{fig:aaflow}). In reality, the agent might additionally carry out further reasoning tasks before creation, including cross-document synthesis or summarization. The LLM generates an output conditioned on the curated context and the original query during \textit{response generation} (Fig.-\ref{fig:aaflow}). 

\begin{figure}[htpb]
    \begin{center}
    \includesvg[width=0.8\linewidth]{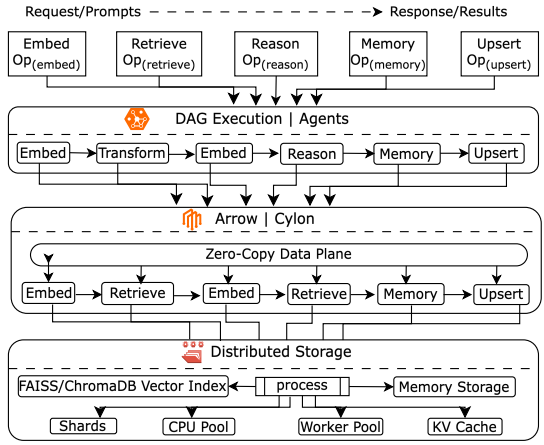}
    \caption{AAFLOW design incorporating a multilayered architecture with agentic DAG execution supported by a zero-copy data plane, used by high-performance agentic execution, hosted in distributed storage, which is managed by a scalable adaptive workflow.}
    \label{fig:agentic_cylon_design}
    \end{center}
\end{figure}

\section{Implementation}

AAFLOW is a constructed distributed runtime over a zero-copy data plane that implements the operator abstraction presented in Section II. The implementation converts each workflow into an execution DAG whose vertices are operator instances and whose edges are typed data dependencies, rather than exposing orchestration as a set of loosely connected framework methods. The runtime then uses explicit communication primitives and batch-aware execution to schedule these operator instances across distant resources and partitioned data.

The implementation consists of four tightly coupled layers: (1) an operator runtime, (2) a zero-copy distributed data plane, (3) a memory-aware retrieval path, and (4) an asynchronous batched execution engine. As illustrated in a layered figure in Fig. \ref{fig:agentic_cylon_design}, these layers together achieve the architectural objectives of operator-driven execution, zero-copy data movement, and resource-deterministic scheduling.

\subsection{Operator Runtime: Realizing the Agentic Abstraction}

AAFLOW materializes each abstract operator,
$Op_{embed}$, $Op_{retrieve}$, $Op_{reason}$, $Op_{memory}$, and $Op_{upsert}$,
as a concrete runtime primitive with explicit input/output schemas, partitioning semantics, and communication behavior.

\textbf{Embedding operator ($Op_{embed}$).}
Tokenization, chunking, and embedding generation are applied independently over each partition by the runtime after it receives document partitions created by preprocessing. A batched map over distributed Arrow \cite{rishabh25} /Cylon \cite{widanage2020high} tables is used to implement this operator. The runtime arranges this stage as embarrassingly parallel work among workers because each chunk is embedded independently. To increase performance while maintaining deterministic partition ownership, embedding models are called using fixed-size micro-batches.

\textbf{Retrieval operator ($Op_{retrieve}$).}
The input query is embedded by the runtime and sent to partition-local vector indices for query-time execution. Every worker compares its shard of the knowledge index or memory index to the local top-$k$ candidates. Local candidates are then globally merged by the runtime to create a final ranked set. Instead of using implicit framework-managed object passing, this stage is implemented as a distributed top-$k$ query with explicit partition-aware routing and reduction.

\textbf{Reasoning operator ($Op_{reason}$).}
A structured reasoning state is created from the recovered context. Before final aggregation, locally acquired evidence is sorted, filtered, and put together into a condensed context representation. AAFLOW generates a deterministic context payload for downstream LLM inference by treating this step as a reduction over recovered fragments. Therefore, reasoning in the implementation is a typed runtime stage that receives ranking evidence and outputs a bounded context object rather than a free-form orchestration callback.

\textbf{Memory operator ($Op_{memory}$).}
Memory is implemented by AAFLOW as a hierarchical state layer that includes long-term vectorized summaries, intermediate reasoning artifacts, and short-term interaction states. Two memory operations are carried out by the runtime: lookup before to reasoning and update following response production. While update employs batched state insertion and summary compaction, lookup follows the same partitioned retrieval path as knowledge search. Instead of treating memory as an ad hoc side channel, this gives it the same execution semantics as other operators.

\textbf{Upsert operator ($Op_{upsert}$).}
Embedding updates generated during ingestion or memory compaction are buffered and written in bulk to partitioned FAISS \cite{johnson2019faiss} indices. In order to minimize write amplification and per-item commit overhead, the runtime groups writes by destination shard and executes batched insertion. One of the primary causes of the observed ingestion benefits is the explicit implementation of this stage as a distributed batched write operator. Other distributed vector databases like ChromaDB \cite{smith2024evaluating, hong2025benchmarking} or Pinecone \cite{pinecone2023} can also be partitioned for bulk indices.

\begin{figure}[htpb]
    \begin{center}
        \includesvg[width=0.9\linewidth]{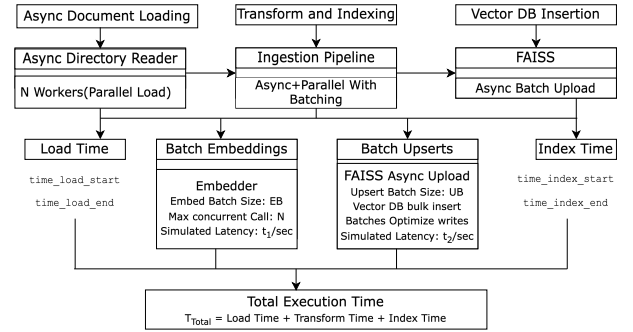}
        \end{center}
        \caption{Advanced Async with Parallel Pipeline of AAFLOW. With batching to measure execution time. BE = Size of the batch embeddings; BU = Size of the upsert batching. $T_{Total} = Load\ Time + Transform\ Time(t_1 + t_2) + Index\ Time$}
    \label{fig:async_pp}
\end{figure}

\subsection{Zero-Copy Data Plane and Distributed Storage}

AAFLOW employs Cylon as the distributed dataframe substrate for partitioned execution and Apache Arrow as the in-memory format for records, chunks, embeddings, and metadata. Between the preprocessing, embedding, retrieval, and indexing phases, this solution eliminates intermediary Python-object serialization. The runtime passes Arrow-backed buffers directly between stages \cite{widanage2020high, abeykoon2022high}, rather than converting across framework-specific containers. This is significant since AAFLOW's architecture aims to eliminate serialization and object-store overhead, two significant bottlenecks in distributed RAG pipelines, in addition to parallelism.

Partitioned tabular datasets with raw text, chunk identifiers, source metadata, and routing information are used to store document collections. Embeddings are attached as columnar vector fields after chunking, and they stay in layouts suitable with Arrow until they are committed to the vector index. To prevent repeated reconstruction during query execution, embeddings are maintained alongside metadata needed for subsequent retrieval, such as provenance, memory scope, and partition location.

FAISS is the main similarity-search backend used by AAFLOW for vector storage. A \emph{knowledge index} for static corpus retrieval and a \emph{memory index} for historical interaction context are the two logical indices that are kept. Because both indices are divided across workers, the same distributed routing architecture may be used for both retrieval and update activities. This implementation detail is crucial because it simplifies scheduling and enhances repeatability by aligning the retrieval and memory operators with the same communication plan.

\subsection{Compiled Workflow Execution}

At runtime, an agentic workflow is first lowered into an execution DAG:
\[
G = (V, E),
\]
where each vertex $v \in V$ corresponds to an instantiated operator and each edge $e \in E$ represents a typed dependency between operator outputs and inputs. The compiler assigns each operator instance to a resource domain based on its execution characteristics:

\begin{itemize}
    \item preprocessing and metadata transforms $\rightarrow$ CPU-distributed partitions,
    \item embedding generation $\rightarrow$ CPU-backed batched workers,
    \item retrieval and merge $\rightarrow$ distributed vector shards + reduction,
    \item reasoning/context assembly $\rightarrow$ bounded aggregation stage,
    \item memory updates and index insertion $\rightarrow$ batched distributed writes.
\end{itemize}

What sets AAFLOW apart from framework-level orchestration is this reduction step. The runtime determines "how" a retrieval or memory action is carried out, but the LLM may determine "what" is required. To put it another way, physical scheduling is not directly determined by logical agent action. Rather, a predetermined operator plan is carried out by the scheduler, guaranteeing stable performance traces, explicit communication, and repeatable stage boundaries.

\subsection{Memory-Aware Retrieval Path}

Retrieval is implemented by AAFLOW as a dual-path query over persistent memory and static knowledge. Upon receiving a question, the runtime creates a query embedding and sends it to the memory index and the knowledge index \cite{karpathy2022agents, intelvector}. A weighted ranking policy over semantic score, source type, and recency is used to merge the resultant candidate sets. As a result, a composite context is created that incorporates both prior interaction state and factual foundation. Three classes of memory artifacts are supported by the implementation: \textbf{persistent long-term memory}: vectorized interaction history or compressed summaries; \textbf{intermediate results}: previously retrieved chunks, partial reasoning outputs, and generated context fragments; and \textbf{agent state}: previous tool calls, decisions, and summaries.

New contexts are selectively promoted into memory by the runtime after generation. While contextual embeddings and reusable summaries are compacted and added to the memory index, short-lived execution traces stay local. Controlling memory development and preventing needless upsert overhead depend on this judicious promotion. Therefore, in terms of implementation, memory is an organized and planned state-management operator rather than a passive cache.

\subsection{Asynchronous Batched Execution Engine}

The key performance-critical portion of the implementation is the asynchronous execution engine used for ingestion and index building, described in Fig. \ref{fig:async_pp}. The pipeline is divided into four distinct steps by AAFLOW: Load, Transform, Embed, and Upsert. The runtime uses stage-local worker pools and constrained queues to connect these instead of considering them as synchronous barriers. Input files are divided and transformed into Arrow/Cylon tables during the \textbf{load stage}. Partition localization and low-overhead ingestion are prioritized in this step. The \textbf{transform stage.} applies chunk creation, normalization, metadata alignment, and delimiter-based splitting across partitions. This CPU-focused stage generates the chunk records that embedding workers use. Chunks are transmitted to specialized embedding workers in fixed-size micro-batches during the \textbf{embed stage}. While separate workers allow overlap with previous and later stages, batching amortizes kernel launch and framework overhead. In the upsert stage, embedding outputs are combined into bigger write batches and added to partitioned FAISS shards. The runtime optimizes index-write efficiency and underlying hardware utilization independently by separating embedding batch size from upsert batch size.

Bounded queues between stages and persistent worker pools are used to implement this asynchronous design. While bounded queues impose backpressure across the pipeline and stop unchecked expansion in the intermediate state, persistent workers avoid recurring startup overhead. In practice, this means that AAFLOW maintains operator-level determinism while overlapping data movement, embedding, and indexing.

\section{Performance Evaluation}

\subsection{Experimental Setup}

Our assessment attempts to evaluate the systems claims presented in Sections I–III. We quantify improvements in preparation, embedding, retrieval, memory access, and index updating stages, which dominate end-to-end execution in distributed RAG systems, rather than trying to speed up LLM decoding itself, in accordance with the scope specified in Section~I.

The \texttt{parallel} queue is used in experiments on the large-scale academic HPC cluster, where each node offers 40 CPU cores. To rigorously isolate data orchestration and framework communication overhead from GPU-bound model computation, we employ a synthetic systems microbenchmark. Rather than using large, instruction-tuned LLMs that would heavily bottleneck the pipeline with raw compute time, we substitute the generation and embedding stages with ultra-lightweight surrogates (distilgpt2 and LocalHashEmbedder). By actively reducing inference time to near-zero, we force the frameworks into an extreme communication-bound regime, exposing their underlying scheduling, zero-copy data plane efficiencies, and baseline systems overhead ($\Omega$). The system architecture remains fully compatible with production-scale neural models. Three scenarios are used to assess AAFLOW. First, we compare end-to-end RAG execution against agentic orchestration frameworks (LangChain~\cite{langchain2023}, LangGraph~\cite{langgraph2024}, CrewAI ~\cite{duan2024exploration, crewai_kickoff_async}, AutoGen~\cite{wu2024autogen}) under equalized concurrency and batching configurations to isolate orchestration overhead. Second, we benchmark ingestion pipelines across multiple distributed baselines (RayDataScalableRAG~\cite{moritz2018ray}, AsyncParallelOnly~\cite{llamaindex_parallel_ingestion}, DaskScalableRAG \cite{rocklin2015dask}, HigressRAG \cite{lin2025higress}) to evaluate stage-wise execution behavior. Third, we evaluate retrieval and response performance, including semantic cache lookup, hybrid retrieval, and non-cached queries. Across all experiments, we measure three system-level properties:
\begin{itemize}
\item \textbf{Pipeline efficiency}: stage-wise latency for Load, Transform, Embed, and Upsert,
\item \textbf{Retrieval performance}: latency and effectiveness of query-time execution,
\item \textbf{Scalability}: behavior under increasing parallelism (strong scaling).
\end{itemize}

Every system makes use of the same hardware configuration, vector storage (FAISS), and embedding models. As a result, execution-model variations are responsible for the observed gains, which include decreased serialization, increased batching efficiency, and less coordination overhead.

\subsection{Framework Benchmarking for RAG Pipeline}

AAFLOW is compared to LangChain, LangGraph, CrewAI, and AutoGen under equal parallelism in Table~\ref{tab:framework-rag-results}. With an end-to-end runtime of 0.8748 seconds, AAFLOW outperforms baseline frameworks, which range from 1.6135 to 1.6447 seconds.

Relative to LangChain, the speedup is:
\[
\frac{1.6447}{0.8748} \approx 1.88\times
\]

\begin{table}[!t]
\caption{RAG Pipeline Benchmark on 32768 tokens generated from 256 documents with Tuned Streaming AAFLOW. All metrics are measured in seconds and TPS = Tokens/s}
\label{tab:framework-rag-results}
\centering
\begin{tabular}{lllllll}
\toprule
Framework  & Load & Trans. & TPS & Embed  & Upsert & Total \\
\midrule
LangChain & 0.0236 & 0.0057 & 94823 & 1.1489 & 0.1403 & 1.6447 \\
LangGraph & 0.0179 & 0.0075 & 93712 & 1.1396 & 0.1364 & 1.6142 \\
CrewAI & 0.0108 & 0.0100 & 93955 & 1.1453 & 0.1326 & 1.6255 \\
AutoGen & 0.0126 & 0.0102 & 94579 & 1.1362 & 0.1352 & 1.6135 \\
AAFLOW & 0.0113 & 0.0115 & 96556 & 0.4856 & 0.0488 & 0.8748 \\
\bottomrule
\end{tabular}
\scriptsize Note: TPS represents the aggregate, cluster-wide throughput across all parallel workers executing the lightweight DistilGPT-2 proxy to saturate the data plane.

\end{table}

Since token throughput is still similar across frameworks, this improvement cannot be attributed to quicker token production. Instead, the gains arise from ingestion-heavy stages:

\begin{itemize}
\item \textbf{Embed}: 0.4856s vs. 1.13–1.15s
\item \textbf{Upsert}: 0.0488s vs. 0.13–0.14s
\end{itemize}

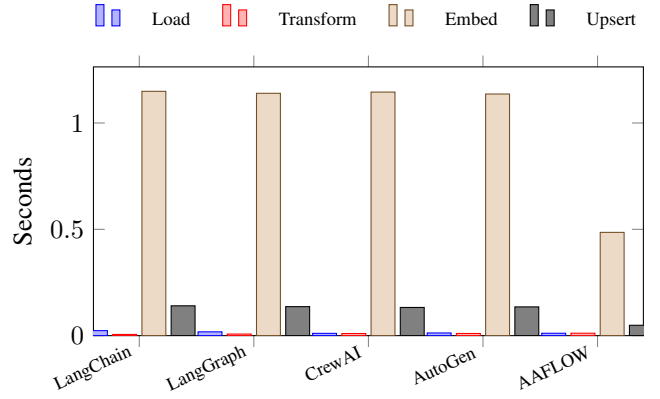
\begin{figure}[!t]
\centering
\begin{tikzpicture}
\begin{axis}[
    ybar,
    bar width=9pt,
    width=\columnwidth,
    height=0.58\columnwidth,
    ylabel={Seconds},
    symbolic x coords={LangChain,LangGraph,CrewAI,AutoGen,AAFLOW},
    xtick=data,
    x tick label style={rotate=25,anchor=east, font=\scriptsize},
    legend style={
        font=\scriptsize,
        legend columns=5,
        at={(0.5,1.1)},
        anchor=south,
        draw=none,
        column sep=0.35cm
    },
    ymin=0,
]
\addplot coordinates {(LangChain,0.0236) (LangGraph,0.0179) (CrewAI,0.0108) (AutoGen,0.0126) (AAFLOW,0.0113)};
\addplot coordinates {(LangChain,0.0057) (LangGraph,0.0075) (CrewAI,0.0100) (AutoGen,0.0102) (AAFLOW,0.0115)};
\addplot coordinates {(LangChain,1.1489) (LangGraph,1.1396) (CrewAI,1.1453) (AutoGen,1.1362) (AAFLOW,0.4856)};
\addplot coordinates {(LangChain,0.1403) (LangGraph,0.1364) (CrewAI,0.1326) (AutoGen,0.1352) (AAFLOW,0.0488)};
\legend{Load,Transform,Embed,Upsert}
\end{axis}
\end{tikzpicture}
\caption{Framework benchmarking in each vital stage in the RAG pipeline under equal parallelism. AAFLOW beats all frameworks in two vital stages (embed and upsert)}
\label{fig:faiss-stage-times}
\end{figure}

The AAFLOW execution model explains these enhancements. In order to minimize per-request overhead ($\alpha$ in Eq.~1), embeddings are first processed in fixed-size micro-batches utilizing persistent workers. Second, write amplification and commit overhead are decreased by grouping upsert operations into bigger batched writes. Third, intermediary serialization between pipeline steps is eliminated via zero-copy data transmission. As illustrated in Fig.~\ref{fig:faiss-stage-times}, token throughput is almost the same and Load and Transform stages are comparable across frameworks. Consequently, Embed and Upsert dominate the performance gap, indicating that end-to-end performance is determined by orchestration efficiency rather than model inference.

\subsection{Hybrid Parallel Ingestion Pipeline Benchmarking}

The ingestion performance of various distributed configurations is assessed in Table~\ref{tab:parallel_benchmarks}. AAFLOW outperforms DaskScalableRAG (16.188s), AsyncParallelOnly (11.641s), and HigressRAG (4.439s) with the lowest overall runtime (3.487s). The Transform, Embed, and Upsert stages show the biggest gains:

\begin{itemize}
\item \textbf{Transform}: 1.277s vs. 2.108s (HigressRAG)
\item \textbf{Embed}: 1.507s vs. 1.765s
\item \textbf{Upsert}: 0.437s vs. 0.634s
\end{itemize}

\begin{table}[t]
\centering
\caption{Small-scale benchmark comparison across parallel configurations (10 million chunks, 4096 files).}
\label{tab:parallel_benchmarks}
\scriptsize
\renewcommand{\arraystretch}{1.1}

\begin{tabular}{
    l
    l
    l
    l
    l
    l
    l
}
\toprule
\textbf{Config}
 & {\textbf{Load}}
 & {\textbf{Trans.}}
 & {\textbf{Embed}}
 & {\textbf{Upsert}}
 & {\textbf{Total}}
 & {\textbf{AAFLOW}} \\
 & {(s)} & {(s)} & {(s)} & {(s)} & {(s)} & {(Boost)} \\
\midrule

RayScalableRAG   & 57.636  & 0.032        & 22.914   & 3.151     & 84.136   & 24.12  \\
AsyncParallelOnly   & 2.646   & 2.047        & 8.067    & 0.724     & 11.641     & 3.33  \\

DaskScalableRAG     & 2.876   & 2.194        & 3.222    & 12.423    & 16.188      & 4.64 \\

HigressRAG     & 0.890   & 2.108        & 1.765    & 0.634     & 4.439 & 1.28 \\

\textbf{AAFLOW}     &  0.952   & 1.277        & 1.507    & 0.437    & \textbf{3.487} & \textbf{1} \\

\bottomrule
\end{tabular}

\vspace{0.4em}
\scriptsize \textit{AAFLOW vs DaskScalableRAG and HigressRAG, total improvement: $4.64\times$ and $1.28\times$ faster with 16 logical workers. Note that AAFLOW's total execution time is less than the sum of its individual stages. This directly validates the effectiveness of our asynchronous pipeline in successfully overlapping stage computation and masking latency.}
\end{table}

These gains follow directly from the execution model in Eq.~(2)--(3). AAFLOW reduces total runtime by:
\begin{itemize}
\item amortizing fixed overhead ($\alpha$) through batching,
\item increasing parallelism ($P$) via persistent workers,
\item minimizing framework overhead ($\Omega$) through zero-copy data exchange and explicit communication.
\end{itemize}

Conversely, RayScalableRAG, Dask and AsyncParallelonly experience increased $\Omega$ as a result of object-store overhead, job scheduling, and serialization. Although HigressRAG uses partial pipeline overlap to save runtime, stage-level synchronization still results in coordination costs.

In the end-to-end pipeline, AAFLOW achieves:
\begin{itemize}
\item \textbf{24.12$\times$ speedup} over RayScalableRAG,
\item \textbf{4.64$\times$ speedup} over DaskScalableRAG,
\item \textbf{1.28$\times$ speedup} over HigressRAG.
\end{itemize}

This indicates that enhancing ingestion performance is mostly achieved by reorganizing the execution model rather than by boosting raw parallelism.

\subsection{Strong and Weak Scaling of Parallel Ingestion}

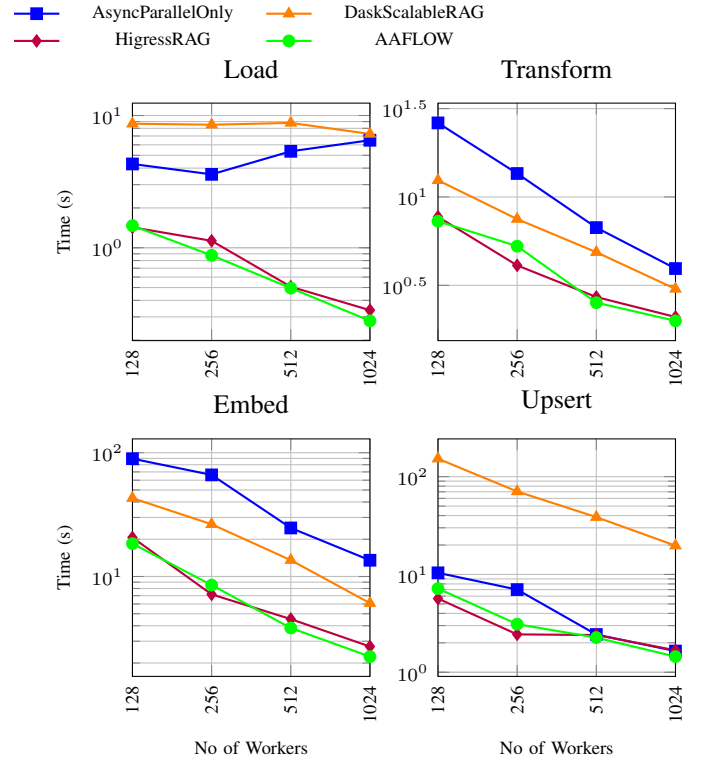
\begin{figure}[t]
\centering
\begin{tikzpicture}
\begin{groupplot}[
  group style = {
    group size = 2 by 2,
    horizontal sep = 0.9cm,
    vertical sep = 1.3cm
  },
  width=0.2\textheight,
  height=0.2\textheight,
  xmode=log,
  log basis x=2,
  xmin=128, xmax=1024,
  xtick={128,256,512,1024},
  xticklabels={128,256,512,1024},
  xticklabel style={font=\scriptsize, rotate=90, anchor=east},
  yticklabel style={font=\scriptsize},
  ylabel style={font=\scriptsize},
  xlabel style={font=\scriptsize},
  grid=both,
  legend cell align={center},
  legend style={
    font=\scriptsize,
    legend columns=2,
    at={(0.5,1.18)},
    anchor=south,
    draw=none,
    column sep=0.35cm
  }
]

\nextgroupplot[
  title={Load},
  ylabel={Time (s)},
  ymode=log
]

\addlegendentry{AsyncParallelOnly}
\addlegendentry{DaskScalableRAG}
\addlegendentry{HigressRAG}
\addlegendentry{AAFLOW}

\addplot[blue, thick,mark=square*] coordinates {
  (128,4.300)(256,3.589)(512,5.362)(1024,6.500)
};

\addplot[orange,thick,mark=triangle*] coordinates {
  (128,8.668)(256,8.518)(512,8.790)(1024,7.252)
};

\addplot[purple,thick,mark=diamond*] coordinates {
  (128,1.427)(256,1.128)(512,0.507)(1024,0.337)
};

\addplot[green, thick,mark=otimes*] coordinates {
  (128,1.469)(256,0.876)(512,0.495)(1024,0.280)
};

\nextgroupplot[
  title={Transform},
  ymode=log
]

\addplot[blue, thick,mark=square*] coordinates {
  (128,26.248)(256,13.577)(512,6.709)(1024,3.936)
};

\addplot[orange,thick,mark=triangle*] coordinates {
  (128,12.429)(256,7.498)(512,4.868)(1024,3.017)
};

\addplot[purple,thick,mark=diamond*] coordinates {
  (128,7.723)(256,4.100)(512,2.715)(1024,2.093)
};

\addplot[green, thick,mark=otimes*] coordinates {
  (128,7.289)(256,5.267)(512,2.523)(1024,1.991)
};

\nextgroupplot[
  title={Embed},
  ylabel={Time (s)},
  xlabel={No of Workers},
  ymode=log
]

\addplot[blue, thick,mark=square*] coordinates {
  (128,89.521)(256,66.224)(512,24.667)(1024,13.528)
};

\addplot[orange,thick,mark=triangle*] coordinates {
  (128,42.996)(256,26.431)(512,13.556)(1024,6.090)
};

\addplot[purple,thick,mark=diamond*] coordinates {
  (128,20.646)(256,7.174)(512,4.526)(1024,2.717)
};

\addplot[green, thick,mark=otimes*] coordinates {
  (128,18.435)(256,8.523)(512,3.838)(1024,2.253)
};

\nextgroupplot[
  title={Upsert},
  xlabel={No of Workers},
  ymode=log
]

\addplot[blue, thick,mark=square*] coordinates {
  (128,10.388)(256,7.005)(512,2.431)(1024,1.647)
};

\addplot[orange,thick,mark=triangle*] coordinates {
  (128,152.708)(256,70.558)(512,38.585)(1024,19.707)
};

\addplot[purple,thick,mark=diamond*] coordinates {
  (128,5.670)(256,2.440)(512,2.402)(1024,1.678)
};

\addplot[green, thick,mark=otimes*] coordinates {
  (128,7.180)(256,3.104)(512,2.261)(1024,1.446)
};

\end{groupplot}
\end{tikzpicture}
\caption{Strong scaling behavior across configurations for Load, Transform, Embed, Upsert operations. 100 million chunks (a 92 GB synthetic corpus generated from \texttt{wikitext2\_train}) are used from 4096 files. Each node has 40 cores, totaling 128--1024 workers are used in multiple runs (e.g., 7 nodes * 40 cores = 280 workers)} 
\label{fig:scaling_all}
\end{figure}

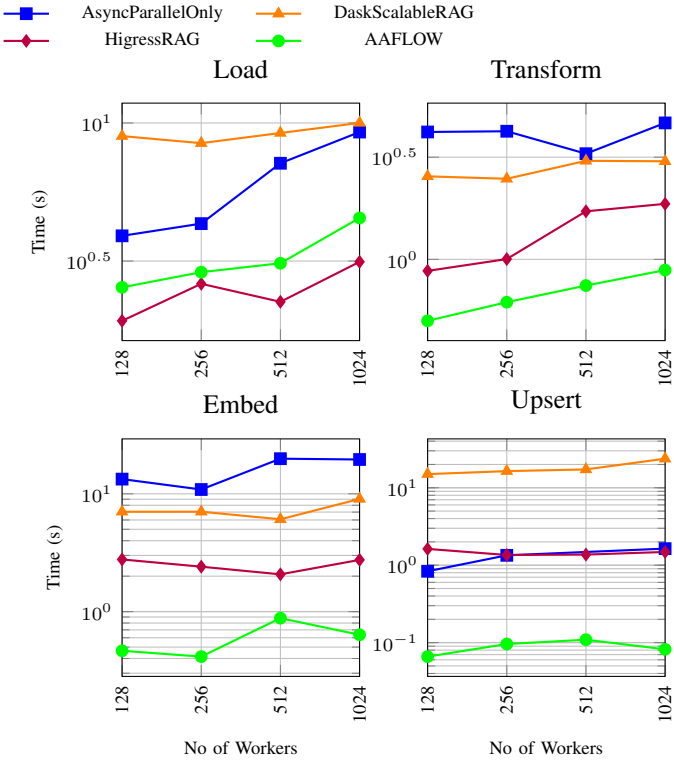
\begin{figure}[t]
\centering
\begin{tikzpicture}
\begin{groupplot}[
  group style = {
    group size = 2 by 2,
    horizontal sep = 0.9cm,
    vertical sep = 1.3cm
  },
  width=0.2\textheight,
  height=0.2\textheight,
  xmode=log,
  log basis x=2,
  xmin=128, xmax=1024,
  xtick={128,256,512,1024},
  xticklabels={128,256,512,1024},
  xticklabel style={font=\scriptsize, rotate=90, anchor=east},
  yticklabel style={font=\scriptsize},
  ylabel style={font=\scriptsize},
  xlabel style={font=\scriptsize},
  grid=both,
  legend cell align={center},
  legend style={
    font=\scriptsize,
    legend columns=2,
    at={(0.5,1.18)},
    anchor=south,
    draw=none,
    column sep=0.35cm
  }
]

\nextgroupplot[
  title={Load},
  ylabel={Time (s)},
  ymode=log
]

\addlegendentry{AsyncParallelOnly}
\addlegendentry{DaskScalableRAG}
\addlegendentry{HigressRAG}
\addlegendentry{AAFLOW}

\addplot[blue, thick,mark=square*] coordinates {
  (128,3.903)(256,4.321)(512,7.148)(1024,9.272)
};

\addplot[orange,thick,mark=triangle*] coordinates {
  (128,8.964)(256,8.452)(512,9.200)(1024,10.006)
};

\addplot[purple,thick,mark=diamond*] coordinates {
  (128,1.921)(256,2.616)(512,2.251)(1024,3.142)
};

\addplot[green, thick,mark=otimes*] coordinates {
  (128,2.540)(256,2.882)(512,3.105)(1024,4.529)
};

\nextgroupplot[
  title={Transform},
  ymode=log
]

\addplot[blue, thick,mark=square*] coordinates {
  (128,4.201)(256,4.236)(512,3.289)(1024,4.653)
};

\addplot[orange,thick,mark=triangle*] coordinates {
  (128,2.548)(256,2.479)(512,3.035)(1024,3.020)
};

\addplot[purple,thick,mark=diamond*] coordinates {
  (128,0.878)(256,1.003)(512,1.718)(1024,1.869)
};

\addplot[green, thick,mark=otimes*] coordinates {
  (128,0.500)(256,0.617)(512,0.744)(1024,0.886)
};

\nextgroupplot[
  title={Embed},
  ylabel={Time (s)},
  xlabel={No of Workers},
  ymode=log
]

\addplot[blue, thick,mark=square*] coordinates {
  (128,13.339)(256,10.883)(512,19.886)(1024,19.568)
};

\addplot[orange,thick,mark=triangle*] coordinates {
  (128,7.047)(256,7.052)(512,6.079)(1024,9.051)
};

\addplot[purple,thick,mark=diamond*] coordinates {
  (128,2.773)(256,2.407)(512,2.073)(1024,2.749)
};

\addplot[green, thick,mark=otimes*] coordinates {
  (128,0.465)(256,0.415)(512,0.880)(1024,0.636)
};

\nextgroupplot[
  title={Upsert},
  xlabel={No of Workers},
  ymode=log
]

\addplot[blue, thick,mark=square*] coordinates {
  (128,0.835)(256,1.341)(1.858)(1024,1.637)
};

\addplot[orange,thick,mark=triangle*] coordinates {
  (128,15.014)(256, 16.398)(512,17.253)(1024,23.742)
};

\addplot[purple,thick,mark=diamond*] coordinates {
  (128,1.623)(256,1.355)(512,1.371)(1024,1.482)
};

\addplot[green, thick,mark=otimes*] coordinates {
  (128,0.066)(256,0.096)(512,0.109)(1024,0.082)
};

\end{groupplot}
\end{tikzpicture}
\caption{Weak scaling behavior across configurations for Load, Transform, Embed, and Upsert operations. 95000 chunks are used by each worker. Each node has 40 cores, totaling 128--1024 workers are used in multiple runs (e.g. 7 nodes * 40 cores = 280 workers)} 
\label{fig:scaling_weak_all}
\end{figure}

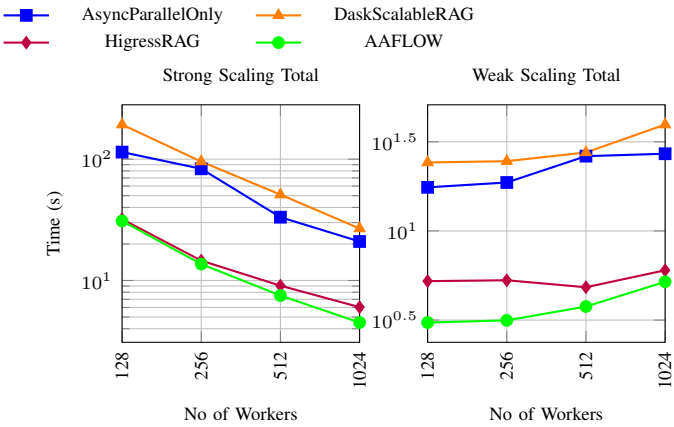
\begin{figure}[t]
\centering

\begin{tikzpicture}
\begin{groupplot}[
  group style = {
    group size = 2 by 1,
    horizontal sep = 0.9cm,
    vertical sep = 1.3cm
  },
  width=0.2\textheight,
  height=0.2\textheight,
  xmode=log,
  log basis x=2,
  xmin=128, xmax=1024,
  xtick={128,256,512,1024},
  xticklabels={128,256,512,1024},
  xticklabel style={font=\scriptsize, rotate=90, anchor=east},
  yticklabel style={font=\scriptsize},
  ylabel style={font=\scriptsize},
  xlabel style={font=\scriptsize},
  title style={font=\scriptsize, yshift=-2pt},
  grid=both,
  legend cell align={center},
  legend style={
    font=\scriptsize,
    legend columns=2,
    at={(0.5,1.18)},
    anchor=south,
    draw=none,
    column sep=0.35cm
  },
  xlabel={No of Workers}
]

\nextgroupplot[
  title={Strong Scaling Total},
  ylabel={Time (s)},
  ymode=log
]

\addlegendentry{AsyncParallelOnly}
\addlegendentry{DaskScalableRAG}
\addlegendentry{HigressRAG}
\addlegendentry{AAFLOW}

\addplot[blue, thick,mark=square*] coordinates {
  (128,114.298)(256,83.266)(512,33.245)(1024,20.955)
};

\addplot[orange,thick,mark=triangle*] coordinates {
  (128,192.660)(256,95.486)(512,51.026)(1024,26.826)
};

\addplot[purple,thick,mark=diamond*] coordinates {
  (128,32.036)(256,14.576)(512,9.063)(1024,6.028)
};

\addplot[green, thick,mark=otimes*] coordinates {
  (128,30.944)(256,13.653)(512,7.514)(1024,4.505)
};

\nextgroupplot[
  title={Weak Scaling Total},
  ymode=log,
]

\addplot[blue, thick,mark=square*] coordinates {
  (128,17.564)(256,18.699)(512,26.276)(1024,27.107)
};

\addplot[orange,thick,mark=triangle*] coordinates {
  (128,24.214)(256,24.627)(512,27.562)(1024,39.506)
};

\addplot[purple,thick,mark=diamond*] coordinates {
  (128,5.228)(256,5.288)(512,4.833)(1024,6.015)
};

\addplot[green, thick,mark=otimes*] coordinates {
  (128,3.064)(256,3.150)(512,3.767)(1024,5.185)
};

\end{groupplot}
\end{tikzpicture}
\caption{Strong and Weak scaling behavior across configurations for Total execution time. 95000 chunks are used by each worker. Each node has 40 cores, totaling 128--1024 workers are used in multiple runs (e.g. 7 nodes * 40 cores = 280 workers)} 
\label{fig:scaling_total_all}
\end{figure}

We evaluate strong and weak scaling behavior by increasing the number of workers from 128 to 1024 (Fig.~\ref{fig:scaling_all} and Fig.~\ref{fig:scaling_weak_all}). AAFLOW reduces end-to-end latency from 30.944s to 4.505s (first graph in Fig.~\ref{fig:scaling_total_all}), achieving over 1.34$\times$ improvement compared to the nearest baseline in strong scaling.
For weak scaling, end-to-end latency experiences an expected moderate increase (from 3.064s to 5.185s, second graph in Fig.~\ref{fig:scaling_total_all}) due to the global reduction overhead ($\Omega$) scaling as a function of worker count. However, it maintains a much shallower degradation curve than baselines. AAFLOW performs ingestion as a non-blocking pipeline with bounded queues and stage-local worker pools, in contrast to conventional frameworks that rely on stage barriers and centralized coordination. This eliminates synchronization overhead and permits overlap between the Load, Transform, Embed, and Upsert stages. Consequently:
\begin{itemize}
\item Load remains approximately constant,
\item Transform scales efficiently with CPU parallelism with minimal communication overhead,
\item Embed benefits from batched CPU execution,
\item Upsert latency decreases due to buffered writes.
\end{itemize}

These findings validate the efficiency of operator-level scheduling and asynchronous batching by demonstrating that AAFLOW sustains near-linear scaling until the bottleneck moves from coordination overhead to intrinsic computation cost.

\subsection{Retrieval and Reasoning Performance}
We compare AAFLOW and HigressRAG \cite{lin2025higress} on retrieval and reasoning operations. HigressRAG is treated as a thinner retrieval path derived from the Higress AI gateway abstraction~\cite{higress_repo}. Table~\ref{tab:faiss-overlap-results} compares AAFLOW and HigressRAG across four scenarios: LLM generation (LLMG), non-cached complex queries (NCCQ), hybrid retrieval (HR), and semantic cache lookup (SCL). AAFLOW consistently reduces total latency:

\begin{itemize}
\item \textbf{LLMG}: 68.23ms $\rightarrow$ 28.12ms (58.8\% reduction)
\item \textbf{NCCQ}: 70.31ms $\rightarrow$ 30.18ms (57.1\% reduction)
\item \textbf{HR}: 21.45ms $\rightarrow$ 1.33ms (93.8\% reduction)
\end{itemize}

\begin{table}[!t]
\caption{Response and Retrieval Benchmark Results with Distributed FAISS.}
\label{tab:faiss-overlap-results}
\centering
\begin{tabular}{lrrrrr}
\toprule
Engine & Scenario  & Retrieval & Mem.  & LLM & Total\\
 &  & {(ms)} & {(ms)} & {(ms)} & {(ms)}\\
\midrule
AAFLOW & LLMG  & 1.48 & 0.02 & 26.53 & 28.12 \\
HigressRAG & LLMG  & 21.55 & 0.00 & 46.61 & 68.23 \\
AAFLOW & NCCQ  & 2.08 & 0.02 & 27.97 & 30.18 \\
HigressRAG & NCCQ  & 22.16 & 0.00 & 48.05 & 70.31 \\
AAFLOW & HR  & 1.26 & 0.02 & 0.00 & 1.33 \\
HigressRAG & HR  & 21.40 & 0.00 & 0.00 & 21.45 \\
AAFLOW & SCL  & 0.00 & 0.00 & 0.00 & 0.03 \\
HigressRAG & SCL  & 0.00 & 0.00 & 0.00 & 0.03 \\
\bottomrule
\end{tabular}

\vspace{0.4em}
\scriptsize \textit{Here, LLM Generation = LLMG, Non-Cached Complex Query = NCCQ, Hybrid Retrieval = HR, Semantic Cache Lookup = SCL, No. of query = 2048}
\end{table}

There are two reasons behind these improvements. In order to minimize unnecessary data transfer, AAFLOW first performs retrieval over partitioned indices utilizing explicit routing and reduction. Second, intermediary conversions during query execution are eliminated by the zero-copy data plane. The 58.8\% reduction in LLMG stage latency is due to the zero-copy data plane eliminating serialization and I/O staging overhead before the LLM engine begins generation, rather than an increase in underlying token generation speed. 
Semantic cache lookup latency is consistent across systems, indicating that execution efficiency, not caching variations, is the source of benefits. These findings show that AAFLOW maintains compatibility with current vector search and generation components while improving retrieval latency and end-to-end response time.

\section{Related Works}

Distributed data systems, workflow runtimes, LLM programming frameworks, LLM serving systems, and memory-augmented retrieval architectures are five closely connected fields of research that are pertinent to AAFLOW.

\textbf{Distributed data systems and dataframe execution:}
Effective preprocessing, partitioning, and communication-aware data transfer are essential for large-scale AI pipelines. While Dask~\cite{rocklin2015dask} offers lightweight task-graph execution for Python-native analytics, Apache Spark~\cite{zaharia2016apache} and Apache Flink~\cite{carbone2015apache} offer general-purpose distributed execution for batch and streaming workloads. Modin provides a crucial precedent for treating high-level abstractions as compilable execution plans rather than framework callbacks~\cite{modin}, arguing that scalable dataframe systems should be based on a clear data model and algebra rather than ad hoc API-level parallelization. By focusing on communication-efficient, Arrow-based distributed execution for HPC and hybrid cloud environments~\cite{widanage2020high,abeykoon2022high,alsaadi2025rhapsody,shan2022hybrid, sarker2022incremental}, Cylon and associated high-performance dataframe work further extend this direction. Building on this system's legacy, AAFLOW focuses on agentic RAG processes rather than just dataframe programming.

\textbf{Workflow runtimes and distributed execution control:}
The scheduling of diverse workloads across dispersed resources is the subject of a second field of study. While ZMQ, Parsl, and RADICAL-Pilot concentrate on task-based workflow execution for HPC and heterogeneous platforms~\cite{babuji2019parsl,merzky2022radical, zmq1:online}, Ray offers a flexible runtime for distributed AI applications~\cite{moritz2018ray}. OneFlow and Google Pathways both investigate asynchronous distributed execution for big machine learning programs~\cite{barham2022pathways, yuan2021oneflow}. An agent-specific operator algebra that maps retrieval, reasoning, memory, and index updates onto communication patterns is not exposed by these systems, despite the fact that they offer crucial execution substrates. In contrast, AAFLOW compiles agentic processes into explicit operator DAGs with zero-copy data transfer and partition-aware scheduling.

\textbf{LLM programming and agent frameworks:}
Multi-step orchestration, tool use, and agent coordination are made possible at the application layer by LangChain~\cite{langchain2023}, LangGraph~\cite{langgraph2024}, CrewAI~\cite{duan2024exploration,crewai_kickoff_async}, and AutoGen~\cite{wu2024autogen}. These frameworks facilitate the creation of agents, although they usually handle execution as a runtime orchestration issue rather than a communication-aware systems issue. By treating LM pipelines as declarative computational graphs and assembling them into optimal LM programs, DSPy advances this trend. This approach is spiritually similar to our operator-centric view of agentic workflows~\cite{dspy}. By reorienting the emphasis from prompt/program specification to distributed execution and data flow, AAFLOW enhances such systems.

\textbf{LLM serving runtimes and inference systems:}
The execution of LLM serving itself is optimized by another significant class of related technologies. By using PagedAttention and effective KV-cache management, vLLM increases serving throughput and shows that careful runtime design may significantly increase model serving efficiency~\cite{vllm}. Through a language/runtime co-design~\cite{sglang}, SGLang also aims to execute structured language-model programs efficiently. Although these systems are quite important, their main focus is on organized LM execution on accelerators and model-serving efficiency. AAFLOW, on the other hand, concentrates on the surrounding distributed pipeline—preprocessing, retrieval, memory access, and index maintenance—where coordination and data movement account for the majority of overall costs.

\textbf{Retrieval-augmented generation and memory systems:}
To enhance factual grounding, classical RAG systems integrate generation and retrieval~\cite{lewis2020retrieval}. This model is expanded with more sophisticated retrieval and memory mechanisms in more recent study. Persistent or structured memory can enhance long-context and multi-step reasoning, as demonstrated by MemoRAG~\cite{qian2025memorag}, RAG-Tuned-LLM~\cite{wei2025tuning}, HippoRAG~\cite{gutierrez2025rag}, and Cue RAG~\cite{fu2025cue}. HigressRAG~\cite{lin2025higress} concentrates on retrieval-path optimization using routing, hybrid retrieval, and semantic cache lookup. AAFLOW complements existing methods by offering a distributed systems architecture where retrieval, memory, and index updates are handled as first-class operators with explicit communication and batching semantics, as opposed to merely introducing a new retrieval heuristic.

In general, current research has improved either the \emph{runtime layer} of distributed systems (task execution, dataframe processing, model serving) or the \emph{application layer} of agentic AI (prompting, orchestration, memory design). By treating agentic workflows as compilable distributed programs, AAFLOW bridges these two viewpoints by integrating resource-deterministic scheduling, zero-copy data transfer, and operator-driven execution into a unified runtime.

\section{Discussion and Future Experiments}

The evaluation results show that rather than LLM inference acceleration, the main performance increases of AAFLOW come from enhancements in data movement, batching efficiency, and execution scheduling. Token throughput is similar across frameworks, as Table I and Table II demonstrate, although there are notable decreases in the embedding and upsert stages. This demonstrates that pipeline orchestration, not model execution, is the primary bottleneck in agentic RAG systems. The findings support the execution model presented in Section II from a systems perspective. The total runtime can be expressed as:
\[
T \approx \frac{N\beta}{P} + \frac{N\alpha}{bP} + \Omega,
\]
where batching reduces per-request overhead ($\alpha$), parallelism reduces the time spent on useful work ($\beta$), and framework overhead ($\Omega$) captures serialization and coordination costs. By decreasing $\Omega$ through zero-copy data exchange and boosting effective batching ($b$) and parallelism ($P$) through persistent workers and asynchronous execution, AAFLOW enhances performance.

A crucial systems insight is further highlighted by the experimental results: greater parallelism by itself does not ensure better performance. Frameworks like AsyncParallel and DaskScalableRAG only show increased overhead as a result of synchronization hurdles, object management, and job scheduling. By reorganizing execution around operator-level scheduling, clear communication patterns, and stage overlap, AAFLOW, on the other hand, delivers more efficiency. This minimizes redundant data transportation, cuts down on idle time, and improves compute and memory resource use. Improvements are centered on retrieval execution and communication-bound stages (Embed and Upsert), which is another significant finding. This is consistent with the architectural design, which minimizes network overhead and eliminates intermediary serialization through partition-aware routing and zero-copy data transfer. The efficiency of asynchronous batching and restricted pipeline execution is demonstrated by the strong and weak scaling findings (Fig. \ref{fig:scaling_all}– \ref{fig:scaling_total_all}), which further confirm that AAFLOW maintains near-linear scaling until computation becomes the major cost.

\subsection{Future Experiments}

Several extensions are required to further characterize system behavior, even if the current evaluation validates the fundamental design ideas. To separate the contribution of each operator (Op$_{embed}$, Op$_{retrieve}$, Op$_{reason}$, Op$_{memory}$, Op$_{upsert}$), we first intend to carry out operator-level micro-benchmarking. This will make it possible to validate communication patterns (broadcast, shuffle, reduce) under different data distributions and to attribute performance gains more precisely. Second, in order to determine how indexing structures and partitioning techniques affect retrieval latency and ingestion performance, we will assess AAFLOW across several vector storage backends (such as FAISS, ChromaDB, and other distributed indices). Third, by contrasting execution with and without the memory operator, we hope to examine the effects of memory integration. This includes measuring the overhead of memory management and compaction as well as evaluating latency, retrieval quality, and system stability under multi-step query workloads. Fourth, in order to precisely measure serialization overhead, network transfer time, and coordination latency across frameworks, we shall do communication-cost profiling. This will improve the causal relationship between system design and observed performance gains by directly measuring the $\Omega$ term in Eq. (3). Lastly, we intend to compare execution traces under identical workloads and analyze variance across repeated runs in order to assess repeatability and execution determinism. In scientific computer contexts, where reliable execution and constant performance are essential, this is especially crucial. The influence of operator-driven execution and zero-copy data movement on large-scale agentic workflows will be better understood thanks to these upcoming experiments.

\section{Conclusion}

The unified distributed runtime AAFLOW, which reformulates agentic RAG workflows as operator-driven execution graphs over high-performance communication primitives, is presented in this paper. AAFLOW bridges the gap between adaptable agentic orchestration and effective distributed systems design by offering an agentic operator algebra, a zero-copy data plane, and a resource-deterministic execution model. Without changing LLM inference, the evaluation shows that AAFLOW achieves significant performance improvements—up to 4.64$\times$ in ingestion pipelines and 1.88$\times$ in end-to-end RAG execution. Rather than modifications to model computation, these benefits result from less serialization overhead, enhanced batching efficiency, and lower coordination costs. AAFLOW ensures effective parallel execution by reducing synchronization costs and permitting stage overlap, as demonstrated by both strong and weak scaling results. This work emphasizes a change in system architecture for AI processes in a broader sense. AAFLOW aggregates agentic pipelines into organized, communication-aware execution plans rather than handling them as dynamic, framework-driven processes. This makes it possible to achieve scalability, repeatability, and predictable performance—all crucial for implementing agentic AI systems in high-performance computing and scientific settings.

AAFLOW shows that changing the surrounding data and execution infrastructure—rather than speeding up model inference—is the key to expanding agentic workflows. AAFLOW offers a framework for creating effective, scalable, and repeatable AI systems by combining data processing, retrieval, memory, and reasoning under a unified execution paradigm.

\bibliographystyle{IEEEtran}
\bibliography{conference_101719}

\end{document}